\documentclass[aps, prl, twocolumn, showpacs, superscriptaddress, floatfix]{revtex4}
\usepackage{natbib}
\usepackage{amsmath} 
\usepackage{amssymb}	
\usepackage{graphicx} 
\usepackage{times}
\usepackage{color}

\begin{document}
\newcommand\bbone{\ensuremath{\mathbbm{1}}}
\newcommand{\ul}{\underline}
\newcommand{\bp}{{\bf p}}
\newcommand{\vl}{v_{_L}}
\newcommand{\vc}{\mathbf}
\newcommand{\be}{\begin{equation}}
\newcommand{\ee}{\end{equation}}
\newcommand{\bk}{{{\bf{k}}}}
\newcommand{\bK}{{{\bf{K}}}}
\newcommand{\cE}{{{\cal E}}}
\newcommand{\bQ}{{{\bf{Q}}}}
\newcommand{\br}{{{\bf{r}}}}
\newcommand{\bg}{{{\bf{g}}}}
\newcommand{\bG}{{{\bf{G}}}}
\newcommand{\hbr}{{\hat{\bf{r}}}}
\newcommand{\bR}{{{\bf{R}}}}
\newcommand{\bq}{{\bf{q}}}
\newcommand{\hx}{{\hat{x}}}
\newcommand{\hy}{{\hat{y}}}
\newcommand{\hd}{{\hat{\delta}}}
\newcommand{\bea}{\begin{eqnarray}}
\newcommand{\eea}{\end{eqnarray}}
\newcommand{\ra}{\rangle}
\newcommand{\la}{\langle}
\renewcommand{\tt}{{\tilde{t}}}
\newcommand{\upa}{\uparrow}
\newcommand{\dna}{\downarrow}
\newcommand{\bS}{{\bf S}}
\newcommand{\vS}{\vec{S}}
\newcommand{\dg}{{\dagger}}
\newcommand{\pdg}{{\phantom\dagger}}
\newcommand{\tphi}{{\tilde\phi}}
\newcommand{\cf}{{\cal F}}
\newcommand{\ca}{{\cal A}}
\renewcommand{\ni}{\noindent}
\newcommand{\ct}{{\cal T}}
\newcommand{\brf}{\bar{F}}
\newcommand{\brg}{\bar{G}}
\newcommand{\jeff}{j_{\rm eff}}


\title{Structural and magnetic properties of the 5$d^2$ double perovskites Sr$_2 B$ReO$_6$ ($B$~$=$~Y, In)}

\author{A. A. Aczel}
\altaffiliation{author to whom correspondences should be addressed: E-mail:[aczelaa@ornl.gov]}
\affiliation{Quantum Condensed Matter Division, Oak Ridge National Laboratory, Oak Ridge, TN 37831, USA}
\author{Z. Zhao}
\affiliation{Materials Science and Technology Division, Oak Ridge National Laboratory, Oak Ridge, TN 37831, USA}
\affiliation{Department of Physics and Astronomy, University of Tennessee, Knoxville, Tennessee 37996, USA}
\author{S. Calder}
\affiliation{Quantum Condensed Matter Division, Oak Ridge National Laboratory, Oak Ridge, TN 37831, USA}
\author{D.T. Adroja}
\affiliation{ISIS Facility, STFC Rutherford Appleton Laboratory, Harwell Oxford, Oxfordshire, OX11 0QX, UK }
\affiliation{Highly Correlated Matter Research Group, Physics Department, University of Johannesburg, P.O. Box 524, Auckland Park 2006, South Africa}
\author{P.J. Baker}
\affiliation{ISIS Facility, STFC Rutherford Appleton Laboratory, Harwell Oxford, Oxfordshire, OX11 0QX, UK }
\author{J.-Q. Yan}
\affiliation{Materials Science and Technology Division, Oak Ridge National Laboratory, Oak Ridge, TN 37831, USA}
\affiliation{Department of Materials Science and Engineering, University of Tennessee, Knoxville, Tennessee 37996, USA}

\date{\today}

\begin{abstract}
We have performed magnetic susceptibility, heat capacity, neutron powder diffraction, and muon spin relaxation experiments to investigate the magnetic ground states of the 5$d^2$ double perovskites Sr$_2$YReO$_6$ and Sr$_2$InReO$_6$. We find that Sr$_2$YReO$_6$ is a spin glass, while Sr$_2$InReO$_6$ hosts a non-magnetic singlet state. By making detailed comparisons with other 5$d^2$ double perovskites, we argue that a delicate interplay between spin-orbit coupling, non-cubic crystal fields, and exchange interactions plays a key role in the great variation of magnetic ground states observed for this family of materials. 
\end{abstract}


\maketitle

\section{I. Introduction}

Transition metal compounds containing heavy 5$d$ atoms are often governed by spin-orbit coupling (SOC), electronic correlations, and crystal field effects of comparable strength \cite{14_witczak}. The relativistic entanglement of correlated orbital and spin degrees of freedom in such materials can drive exotic quantum states of matter, such as effective $J$~$=$~1/2 Mott insulators \cite{08_kim}, topological superconductors and insulators \cite{11_wang, 10_yang, 10_pesin, 12_carter}, and quantum spin liquids \cite{09_jackeli}. The research community has placed a great deal of emphasis on 5$d^5$ systems due to the extreme quantum nature of the novel effective $J$~$=$~1/2 state associated with that particular $d$-electron configuration. On the other hand, far less work has focused on 5$d^1$-5$d^4$ materials even though strong SOC may play a key role in determining their magnetic properties. 

Double perovskites (DPs) of the form $A_2BB'$O$_6$, with magnetic 4$d$ or 5$d$ $B'$ ions, provide an outstanding opportunity to investigate the interplay of SOC, structural distortions, and $d$-electron configuration on the magnetism of the geometrically-frustrated face-centered-cubic (fcc) lattice \cite{15_vasala}. Hundreds of DPs have now been synthesized in polycrystalline form, and subsequent characterization work revealed a diverse array of magnetic properties. The $5d^5$ DPs that have been studied in detail, including La$_2$MgIrO$_6$ and La$_2$ZnIrO$_6$, are effective $J$~$=$~1/2 candidates \cite{13_cao} that exhibit antiferromagnetic ground states and interesting magnetic excitations \cite{15_cook, 15_aczel}. Materials composed of magnetic ions with 4$d^3$/5$d^3$ electron configurations generally host ordered magnetic ground states also \cite{83_battle, 84_battle, 89_battle, 03_battle, 15_kermarrec, 15_taylor}. Aside from a few known exceptions \cite{13_aczel, 14_aczel}, the spin arrangements in $d^3$ systems are usually well-accounted for by mean field theory, with ground state selection arising from the signs and magnitudes of the nearest neighbor and next nearest neighbor exchange interactions \cite{01_lefmann}. On the other hand, the magnetic order of 4$d^3$/5$d^3$ and 5$d^5$ systems often gives way to exotic magnetic states in 4$d^1$/5$d^1$ quantum spin DPs. For example, Ba$_2$YMoO$_6$ has been described as a valence bond glass with a collective spin-singlet ground state \cite{10_devries, 10_aharen} and Sr$_2 B$ReO$_6$ ($B$~$=$~Mg, Ca) are best characterized as unconventional spin glasses, since no obvious source of disorder has been found \cite{02_wiebe, 03_wiebe, 11_greedan}. 

The magnetism of heavy metal $d^2$ DPs serves as an interesting intermediate case between the $d^1$ and $d^3$/$d^5$ analogues, and recent theoretical work has established a phase diagram for the cubic limit \cite{11_chen}. While there are currently no experimental examples of 4$d^2$ DPs, previous work on 5$d^2$ systems has found evidence for both long-range ordered and alternative ground states. For example, a low-moment, long-range antiferromagnetic ground state has been reported for Ba$_2$CaOsO$_6$ \cite{14_thompson}, spin freezing without signs of long-range magnetic order (i.e. spin glass) has been found for Ca$_2$MgOsO$_6$ \cite{15_yuan} and Ba$_2$YReO$_6$ \cite{10_aharen_2}, and a non-magnetic singlet ground state was proposed for SrLaMgReO$_6$ \cite{15_thompson}. There are also many other 5$d^2$ DPs that have not been characterized in detail. For example, there is no work on the magnetism of Sr$_2$YReO$_6$, and only one report on Sr$_2$InReO$_6$ that presents extremely limited information about its magnetic properties \cite{11_gao}. 

There is currently no consensus on the factors that contribute to ground state selection or their order of importance for 5$d^2$ DPs. In this work, we make progress towards addressing this issue. We first revisit the crystal structures of the largely unexplored, 5$d^2$ monoclinic DPs Sr$_2$YReO$_6$ and Sr$_2$InReO$_6$ with neutron powder diffraction and find that both systems crystallize in the {\it P$2_1$/n} space group with minimal $B/B'$ site mixing. We then determine the magnetic ground states of these systems with a combination of magnetic susceptibility, heat capacity, neutron powder diffraction, and muon spin relaxation. Our combined measurements suggest that Sr$_2$YReO$_6$ is a spin glass, while Sr$_2$InReO$_6$ hosts a non-magnetic singlet ground state. By making extensive comparisons between Sr$_2$YReO$_6$, Sr$_2$InReO$_6$, other 5$d^2$ DPs, and Ni$^{2+}$ $S$~$=$~1 DPs, we propose that the diverse magnetic ground states of the 5$d^2$ family arise from the combined effects of enhanced SOC, exchange interactions, and the great sensitivity of this particular $d$ electron configuration to even modest structural distortions.   

\section{II. Experimental Details}

Polycrystalline Sr$_2B$ReO$_6$ ($B$~$=$~Y and In) samples were synthesized by a conventional solid state reaction method. The starting materials of SrO, Y$_2$O$_3$ (or In$_2$O$_3$), Re and Re$_2$O$_7$ powders were weighed in an appropriate ratio and mixed thoroughly inside of a dry He glovebox. The homogeneous mixture was pelletized and transfered to a 5~ml Al$_2$O$_3$ crucible. The Al$_2$O$_3$ crucible was then sealed in a vacuum quartz tube backfilled with 1/3 atmosphere of high purity Ar. The ampoule was fired in a box furnace at 900$^\circ$C for 48 hours. The fired pellets were reground, repelletized and then fired in a sealed quartz tube for another 100 hours. Room temperature x-ray powder diffraction was performed on a PANalytical X'Pert Pro MPD powder x-ray diffractometer using Cu K$_{\alpha1}$ radiation and confirmed that the samples were of single phase. The DC magnetic susceptibilities were measured between 2-350 K using a Quantum Design (QD) magnetic property measurement system. The temperature dependence of the specific heat was measured between 2-200 K using a QD physical property measurement system. 

Neutron powder diffraction (NPD) experiments were performed with 10~g of Sr$_2$YReO$_6$ and 10~g of Sr$_2$InReO$_6$ at Oak Ridge National Laboratory using the HB-2A powder diffractometer of the High Flux Isotope Reactor. Data was collected with neutron wavelengths of $\lambda$~$=$~1.54~\AA~with collimation 12'-21'-12' and $\lambda$~$=$~2.41~\AA~with collimation 12'-open-12'. The shorter wavelength provides greater intensity and higher $Q$ coverage that was used to investigate the crystal structures, while the longer wavelength provides the lower $Q$ coverage required for magnetic diffraction. The $\mu$SR experiments were performed at the ISIS Pulsed Neutron and Muon Source, UK. We used the EMU and MuSR spectrometers in longitudinal geometry for investigating Sr$_2$InReO$_6$ (closed cycle refrigerator, base $T$~$=$~7.5~K) and Sr$_2$YReO$_6$ (He-flow cryostat, base $T$~$=$~1.4~K) respectively. The time-evolution of the spin polarization of the muon ensemble was measured via the asymmetry function, $A(t)$ \cite{97_dereotier}.   

\section{III. Results}

\subsection{A. Crystal Structures}

\begin{figure}
\centering
\scalebox{0.58}{\includegraphics{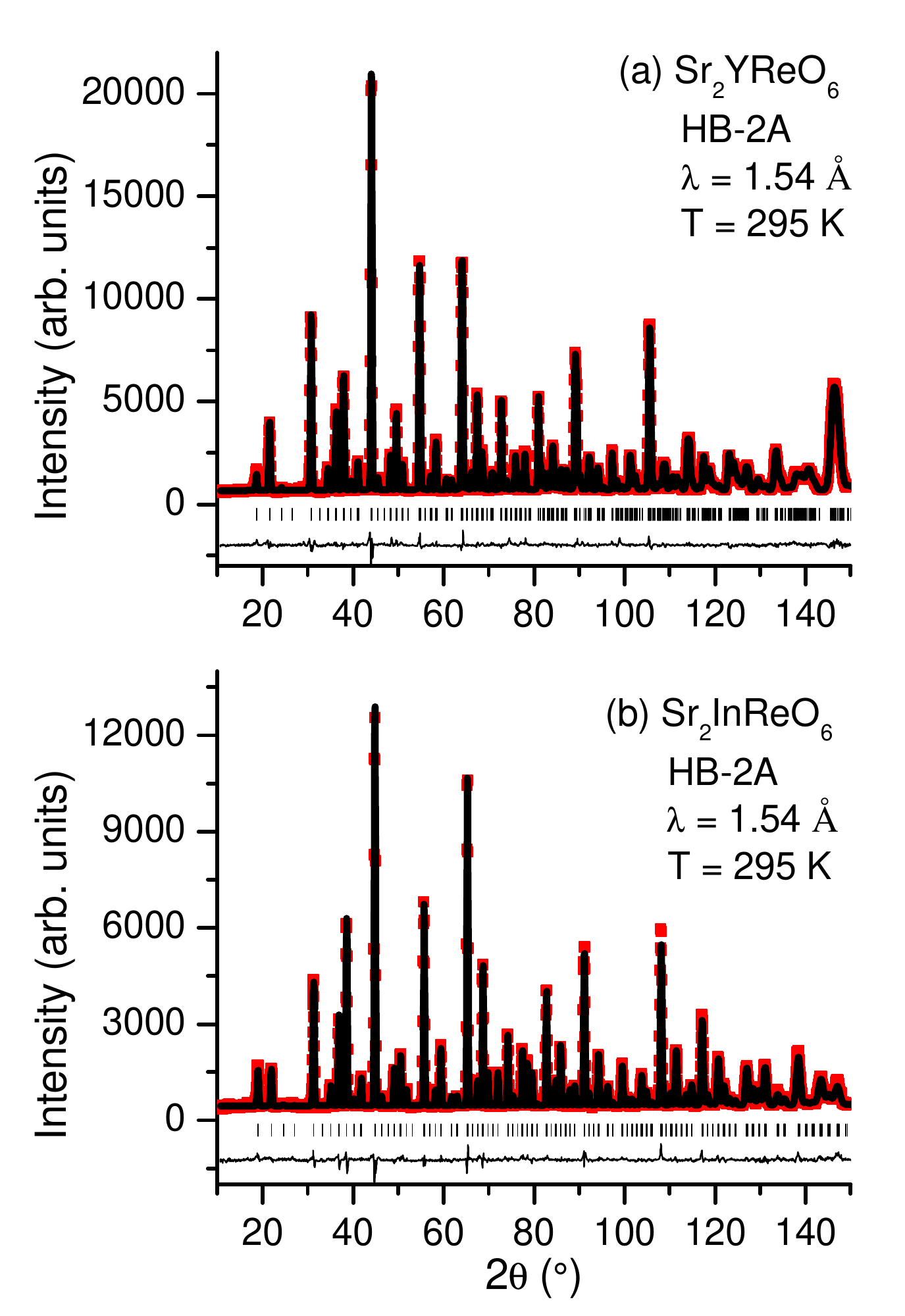}}
\caption{\label{Fig1} Neutron powder diffraction measurements with a wavelength of 1.54~\AA~at $T$~$=$~295~K for both (a) Sr$_2$YReO$_6$ and (b) Sr$_2$InReO$_6$.}
\end{figure}

Figure~\ref{Fig1} shows NPD data collected using $\lambda$~$=$~1.54~\AA~at $T$~$=$~295~K for Sr$_2$YReO$_6$ and Sr$_2$InReO$_6$. Rietveld refinements were performed using FullProf \cite{93_rodriguez}; Table I shows lattice constants and atomic fractional coordinates extracted from the NPD data, while Table II presents selected bond distances and angles. There is very little difference in any of the structural parameters between $T$~$=$~295 and 4~K, so only the low temperature values are shown in most cases. 

\begin{center}
\begin{table}[htb]
\caption{Lattice constants and atomic fractional coordinates for Sr$_2$YReO$_6$ and Sr$_2$InReO$_6$ extracted from the refinements of the $\lambda$~$=$~1.54~\AA~neutron powder diffraction data. } 

\medskip

(a) Sr$_2$YReO$_6$ \\
Space group {\it P2$_1$/n} \\ 

\begin{tabular}{| c | c | c |}
\hline 
\multicolumn{3}{|c|}{Lattice parameters and refinement quality} \\
\hline
T & 4~K & 295~K \\  \hline
a & 5.7779(1)~\AA & 5.7892(1)~\AA \\
b & 5.8193(1)~\AA & 5.8123(1)~\AA \\ 
c & 8.1698(1)~\AA & 8.1970(1)~\AA \\ 
$\beta$ & 90.342(1)$^\circ$ & 90.214(1)$^\circ$ \\
R$_{wp}$ & 4.74~\% & 5.30~\% \\ \hline
\end{tabular}

\medskip

\begin{tabular}{| c | c | c | c | c |}
\hline
\multicolumn{5}{|c|}{Atom positions at T~$=$~4~K} \\
\hline
Atom & Site & x & y & z  \\  \hline
Sr & 4e & 0.5075(4) & 0.5328(2) & 0.2493(3)    \\  
Y & 2c & 0 & 0.5 & 0  \\  
Re & 2d & 0.5 & 0 & 0  \\  
O$_1$ & 4e & 0.3021(4) & 0.7277(4) & 0.9594(3) \\  
O$_2$ & 4e & 0.4272(4) & 0.9826(3) & 0.2329(3) \\
O$_3$ & 4e & 0.2318(4) & 0.1986(4) & 0.9619(3) \\ \hline    
\end{tabular}

\medskip

(b) Sr$_2$InReO$_6$ \\
Space group {\it P2$_1$/n} \\ 

\begin{tabular}{| c | c | c |}
\hline
\multicolumn{3}{|c|}{Lattice parameters and refinement quality} \\
\hline
T & 4~K & 295~K \\ \hline
a & 5.6894(1)~\AA & 5.7045(1)~\AA \\
b & 5.7131(1)~\AA & 5.7059(1)~\AA \\ 
c & 8.0352(1)~\AA & 8.0630(1)~\AA \\ 
$\beta$ & 89.809(1)$^\circ$ & 89.922(2)$^\circ$ \\
R$_{wp}$ & 6.64~\% & 6.13~\% \\ \hline
\end{tabular}

\medskip

\begin{tabular}{| c | c | c | c | c |}
\hline 
\multicolumn{5}{|c|}{Atom positions at T~$=$~4~K} \\
\hline
Atom & Site & x & y & z  \\  \hline
Sr & 4e & 0.4934(7) & 0.5281(3) & 0.2502(6)    \\  
In & 2c & 0 & 0.5 & 0  \\  
Re & 2d & 0.5 & 0 & 0  \\  
O$_1$ & 4e & 0.2972(6) & 0.2827(6) & 0.0385(5) \\  
O$_2$ & 4e & 0.2309(6) & 0.7872(1) & 0.0311(5) \\  
O$_3$ & 4e & 0.5699(5) & 0.9871(4) & 0.2365(4) \\  \hline
\end{tabular}

\end{table}
\end{center}

\begin{center}
\begin{table}[htb]
\caption{Selected bond distances (\AA) and angles ($^\circ$) for Sr$_2$YReO$_6$ and Sr$_2$InReO$_6$ at $T$~$=$~4~K extracted from the refinements of the $\lambda$~$=$~1.54~\AA~neutron powder diffraction data. } 

\begin{tabular}{| c | c | c |}
\hline 
 & Sr$_2$YReO$_6$ & Sr$_2$InReO$_6$ \\  \hline
Re-O$_1$ & 1.981(2) & 2.008(3) \\
Re-O$_2$ & 1.954(2) & 1.970(3) \\ 
Re-O$_3$ & 1.957(2) & 1.944(3) \\ 
Re-O$_1$-$B$ & 155.07(9) & 154.97(13) \\
Re-O$_2$-$B$ & 155.98(10) & 160.98(10) \\ 
Re-O$_3$-$B$ & 156.67(9) & 157.17(13) \\ \hline
\end{tabular}

\end{table}
\end{center}

A previous report \cite{69_baud} proposed that Sr$_2$YReO$_6$ crystallizes at room temperature in the cubic space group {\it Fm$\bar{3}$m}, but there are extra reflections in the NPD data that cannot be described by such a high crystal symmetry. We instead find that the data refines best in the {\it P2$_1$/n} space group; this low crystal symmetry is common for many DPs. Sr$_2$InReO$_6$ was originally characterized as a {\it Fm$\bar{3}$m} system \cite{62_sleight} at room temperature also. However, recent work revisited the crystal structure and revised the room temperature space group to {\it P2$_1$/n} \cite{11_gao}, which is in good agreement with the current results. We find no evidence for a structural phase transition between 295 and 4~K in either material. The tolerance factor $t$ is often used to characterize structural distortions in DPs, and it is defined as $t$~$=$~($r_A + r_O$)/$(\sqrt{2}(r_{<B,B'>} + r_O))$ \cite{15_vasala}, where $r_A$ and $r_O$ are the ionic radii of $A$ and O respectively, and $r_{<B,B'>}$ is the average ionic radius of $B$ and $B'$. DPs with $t$~$=$~1 or slightly greater are generally cubic, while the crystal symmetry decreases with decreasing $t$. Using values of ionic radii from Ref.~\cite{76_shannon}, we find that Sr$_2$YReO$_6$ and Sr$_2$InReO$_6$ have tolerance factors of 0.94 and 0.97 respectively; these values are consistent with their monoclinic crystal symmetries. Finally, the Rietveld refinements confirm that there is essentially no site mixing between the Y and Re atomic positions for Sr$_2$YReO$_6$, while $\sim$~5(1)\%~excess Re was found on the In site for Sr$_2$InReO$_6$. These results are consistent with the larger ionic radii difference between Y (0.9~\AA) and Re (0.58~\AA), as compared to In (0.8~\AA) and Re \cite{76_shannon}.  

\subsection{B. Sr$_2$YReO$_6$: Magnetic Properties}

\begin{figure*}
\centering
\scalebox{0.47}{\includegraphics{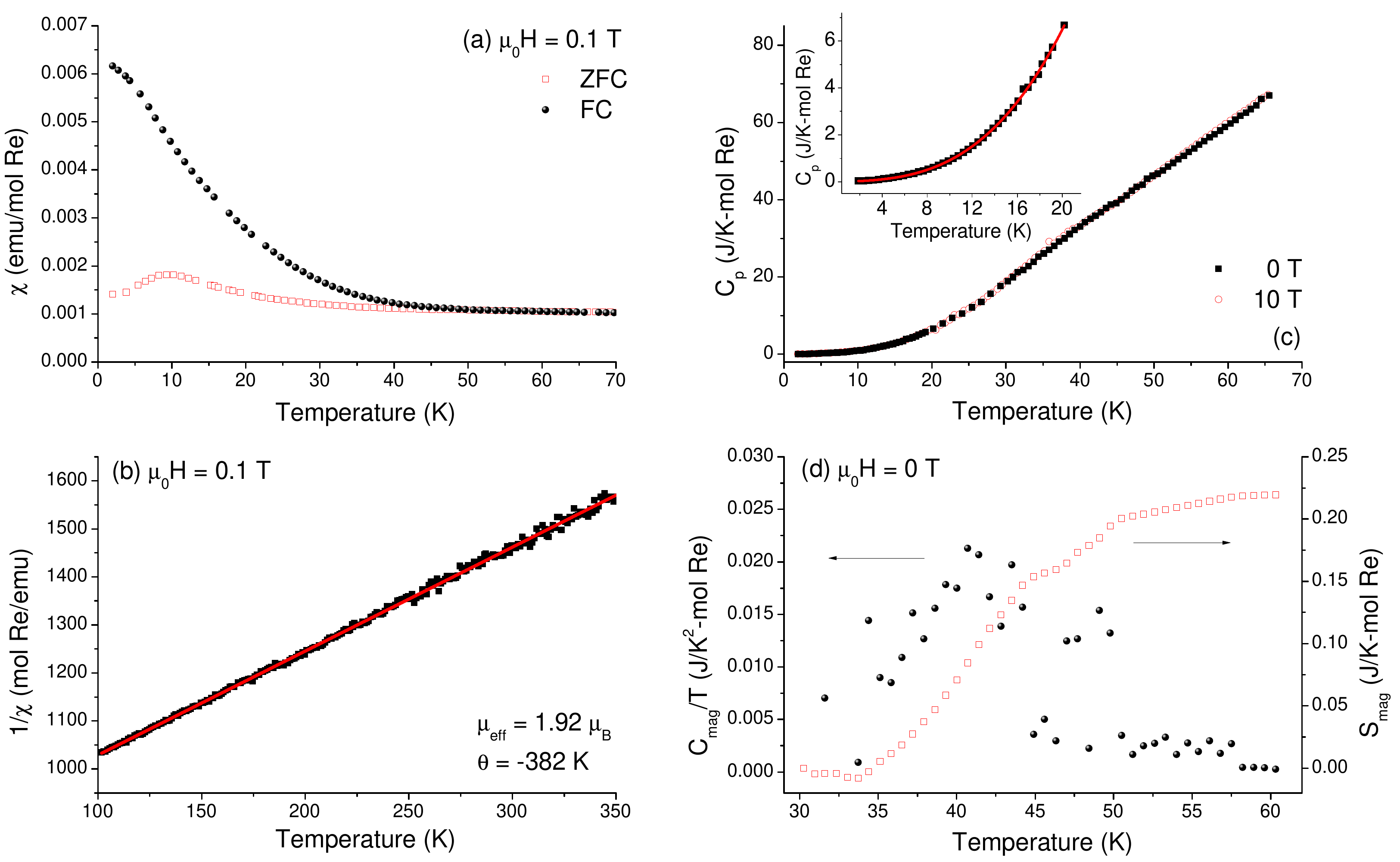}}
\caption{\label{Fig2} (a) DC magnetic susceptibility $\chi$ of Sr$_2$YReO$_6$ measured in an applied field $\mu_0 H$~$=$~0.1~T. Data collected under zero-field-cooled (ZFC) and field-cooled (FC) conditions diverges below 50~K. (b) The $\chi$ data fits well to a Curie-Weiss law plus a constant for 100~$\le$~$T$~$\le$~350~K, with an effective moment $\mu_{eff}$~$=$~1.92~$\mu_B$ and a Curie-Weiss temperature of -382~K. (c) The heat capacity $C_p$ of Sr$_2$YReO$_6$ for both $\mu_0 H$~$=$~0 and 10~T. A weak, broad feature is visible in the data around 40~K. The inset shows the low-$T$ $C_p$ data in zero field, with the fit described in the text illustrated by a solid curve. (d) The magnetic contribution $C_{mag}$ to $C_p$ over the $T$-range 30-60~K. The entropy $S_{mag}$ released over this entropy range is also indicated.  }
\end{figure*}

Figure~\ref{Fig2}(a) and (b) depict the DC magnetic susceptibility measured in an applied field of $\mu_0 H$~$=$~0.1~T, with $\chi$~$=$~$M/\mu_0 H$ (magnetization/applied field). The high temperature data is well-described by a Curie-Weiss law plus a constant, with a fit between 100 and 350~K yielding $\theta_{CW}$~$=$~-382~K and an effective moment $\mu_{eff}$~$=$~1.92~$\mu_B$. The large, negative Curie-Weiss temperature is indicative of strong, dominant antiferromagnetic interactions. The magnitude of $\mu_{eff}$ is reduced from the $S$~$=$~1 spin-only value of 2.83~$\mu_B$, but consistent with reported values for other Re$^{5+}$ DPs \cite{10_aharen_2}. There are deviations from the Curie-Weiss law at low $T$; the zero-field-cooled (ZFC) and field-cooled (FC) datasets begin to diverge at 50~K, and the former also has a broad maximum at 12~K. Overall, the temperature-dependence of $\chi$ is reminiscent of previous results on the spin glass Sr$_2$CaReO$_6$ \cite{02_wiebe}. For that system, the temperature corresponding to the broad maximum in the ZFC $\chi$ data was defined as the freezing temperature $T_f$; therefore we assume that $T_f$~$=$~12~K in the present case. This assigned freezing temperature is supported by our $\mu$SR results presented later.

The heat capacity $C_p$ of Sr$_2$YReO$_6$ is shown in Fig.~\ref{Fig2}(c) in fields of both 0 and 10 T. There is no clear $\lambda$ anomaly indicative of long-range magnetic ordering and essentially no field dependence. The low temperature data ($T$~$\le$~20~K) fits well to the function $C_p$~$=$~$\gamma T + \beta T^3$ with $\gamma$~$=$~15.0(1)~mJ/K$^2$-mol Re and $\beta$~$=$~0.774(5)~mJ/K$^4$-mol Re. A non-zero $\gamma$ is unexpected in insulators that exhibit magnetic order at low $T$, and this feature has been interpreted as a signature of spin glass behavior in insulating Sr$_2$MgReO$_6$ \cite{03_wiebe} and Li$_4$MgReO$_6$ \cite{00_bieringer}. Therefore, we assume that the linear component represents the magnetic contribution $C_{mag}$. Integrating $C_{mag}$/$T$ up to the freezing temperature $T_f$~$=$~12~K yields an entropy of 0.18~J/K-mol Re. Although this value represents only 2\% of the expected entropy release for an $S$~$=$~1 spin glass, the drastically-reduced value is consistent with reports of spin glassiness in Sr$_2$MgReO$_6$ (3\% entropy release at $T_f$) \cite{03_wiebe}, Li$_4$MgReO$_6$ (14\% entropy release below $T_f$) \cite{00_bieringer} and the jarosite (H$_3$O)Fe$_3$(SO$_4$)$_2$(OH)$_6$ (6\% entropy release at $T_f$) \cite{98_wills}. Furthermore, it is well-known that many spin glasses lose significant entropy above $T_f$ \cite{93_mydosh}. 

\begin{figure}
\centering
\scalebox{0.32}{\includegraphics{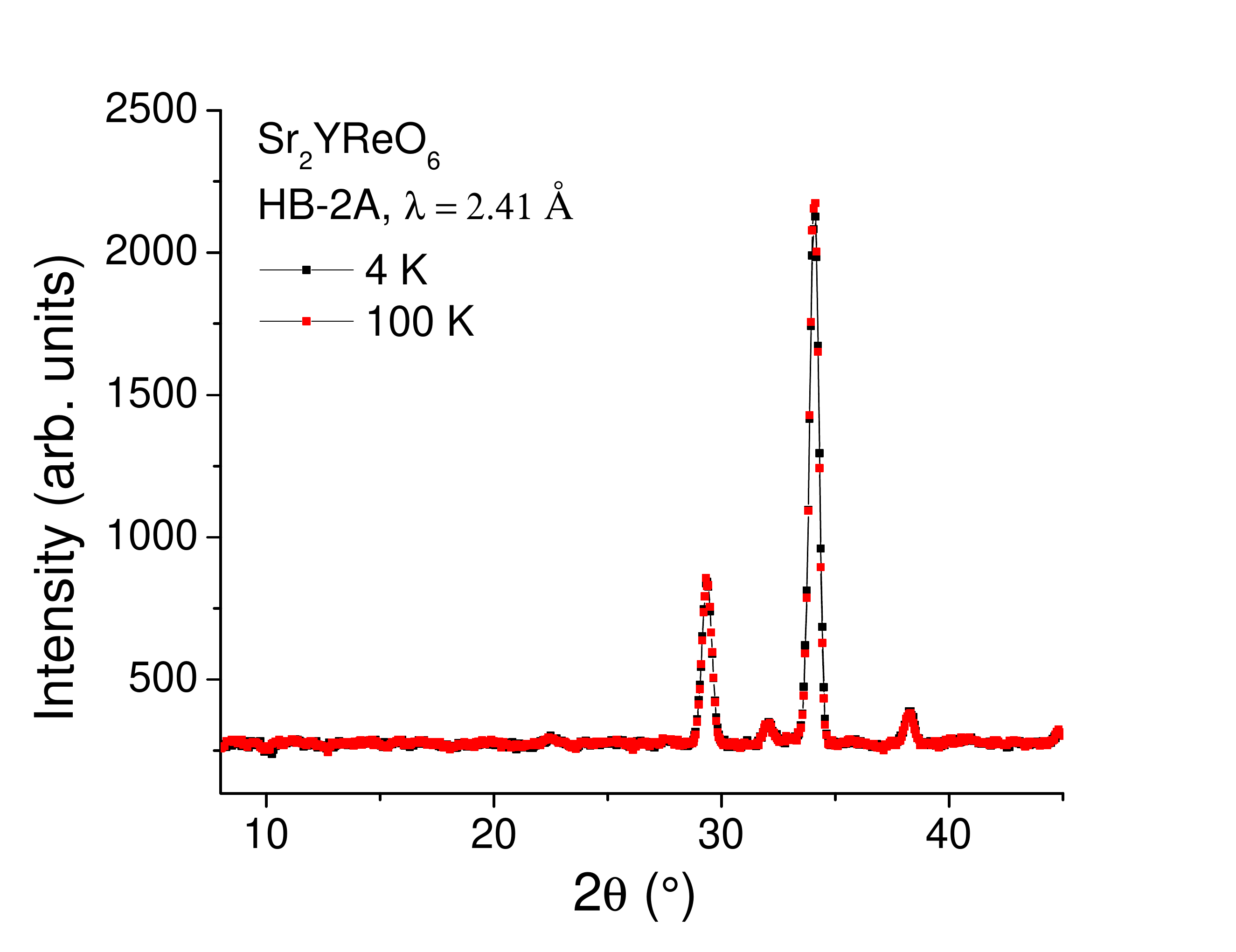}}
\caption{\label{Fig3} Neutron powder diffraction measurements with a wavelength of 2.41~\AA~at $T$~$=$~4 and 100~K for Sr$_2$YReO$_6$. }
\end{figure}

An extremely weak and broad feature is visible in the higher temperature $C_p$ data around 40~K, as shown in Fig.~\ref{Fig2}(c). This feature was assumed to be an additional magnetic component $C_{mag}$, and it was isolated by fitting the data between 30 and 60~K to a polynomial function and then subtracting off the fit as background. The result, plotted as $C_{mag}$/$T$, is shown in Fig.~\ref{Fig2}(d). There is a significant $C_{mag}$ contribution over the $T$ range 35-50~K, which corresponds well to the onset temperature of the ZFC/FC divergence in $\chi$. These combined features are likely indicative of the presence of short-range magnetic correlations at temperatures higher than $T_f$. We note that the entropy released through the broad heat capacity anomaly is 0.22~J/K-mol Re, which is only an additional 2.5\% of the total entropy release expected for an $S$~$=$~1 spin glass. It is possible that additional entropy is released over a greater $T$ range, but no other broad features are apparent in the raw $C_p$ data.  

\begin{figure*}
\centering
\scalebox{0.46}{\includegraphics{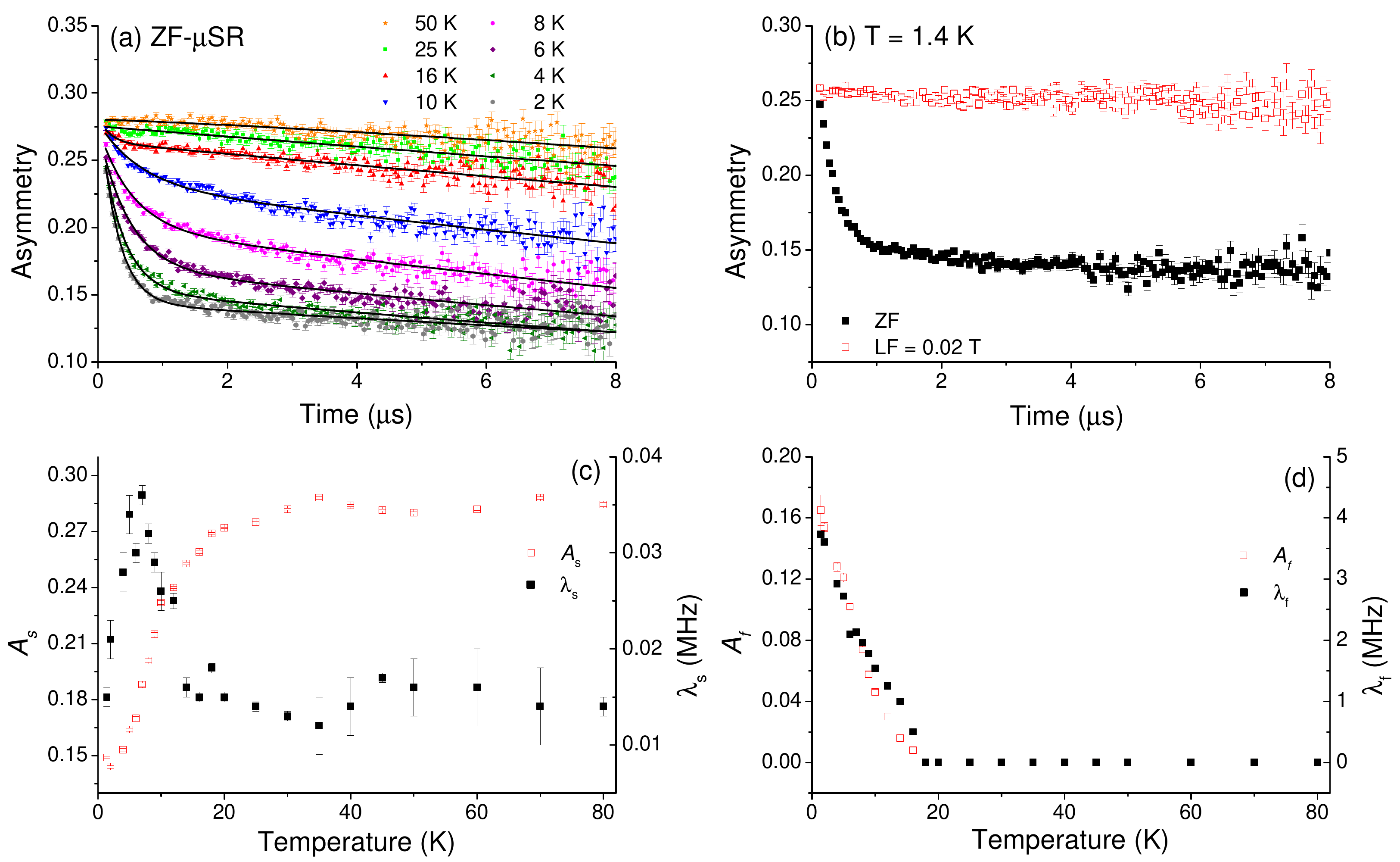}}
\caption{\label{Fig4} (a) Zero field (ZF)-$\mu$SR spectra for Sr$_2$YReO$_6$ at selected temperatures. (b) ZF and longitudinal field (LF)-$\mu$SR results at $T$~$=$~1.4~K. The ZF relaxation associated with the electron spins is completely decoupled in a modest field of 0.02~T, which suggests that it has a static origin. (c) $T$-dependence of the asymmetry and relaxation rate for the slow-relaxing component of the $\mu$SR spectra. (d) $T$-dependence of the asymmetry and relaxation rate for the fast-relaxing component of the $\mu$SR spectra.}
\end{figure*}

Neutron powder diffraction data measured with a wavelength of 2.41~\AA~for Sr$_2$YReO$_6$ is depicted in Fig.~\ref{Fig3} with $T$~$=$~4 and 100~K. The two diffraction patterns are nearly identical, and therefore there is no evidence for long-range magnetic order in this data. This result is in sharp contrast to NPD results on $S$~$=$~1 Ni$^{2+}$ DPs, which always find evidence for a long-range ordered magnetic ground state with an ordered moment between 1.5-2.2~$\mu_B$/Ni$^{2+}$ \cite{67_cox, 02_rodriguez, 92_attfield, 08_chakraborty}. The null diffraction result for Sr$_2$YReO$_6$ implies that the ordered Re moment is drastically-reduced as compared to its Ni counterparts or long-range order is absent down to 4~K. The former is possible due to increased covalency or enhanced spin-orbit coupling, and has been experimentally proven for Ba$_2$CaOsO$_6$ \cite{14_thompson}, where an ordered moment of $\sim$~0.2$\mu_B$ was estimated from muon spin relaxation ($\mu$SR) measurements. 

The $\mu$SR technique \cite{97_dereotier} allows additional constraints to be placed on the possible magnetic ground state for Sr$_2$YReO$_6$. $\mu$SR is an extremely sensitive probe of magnetism and can readily differentiate between ordered, glassy, and dynamic ground states; ordered moments as small as 0.01~$\mu_B$ can be easily detected. Figure~\ref{Fig4} summarizes $\mu$SR results collected with the MuSR spectrometer at ISIS. Figure~\ref{Fig4}(a) shows asymmetry spectra at selected temperatures measured in zero applied field (ZF). No missing asymmetry or spontaneous oscillations, as would be expected for a long-range ordered ground state, are apparent in the data. Instead, the data features two relaxing components and therefore was fit to the following function:
\begin{equation}
A(t) = A_s e^{-(\lambda_s t)^\beta} + A_f e^{-\lambda_f t} 
\end{equation}
where the first and second terms represent the slow and fast-relaxing components of the spectra respectively. At the highest temperatures measured, only the first term was required in the fitting. Furthermore, $\beta$ was found to decrease from $\sim$~1.5 at $T$~$=$~80~K to 1 at $T$~$=$~30~K, and therefore was subsequently fixed to 1 for $T$~$<$~30~K. The decrease in $\beta$ at lower temperatures is consistent with the slowing down of electron spins into the $\mu$SR time window; Gaussian relaxation due to static nuclear moments dominates at high temperature.  

The fast component of the signal begins to develop just below 20~K as shown in Fig.~\ref{Fig4}(d); the relaxation rate $\lambda_f$ and the amplitude $A_f$ continue to increase down to 1.4 K. The $T$-dependence of the relaxation rate $\lambda_s$ for the slow component depicted in Fig.~\ref{Fig4}(c) is drastically different, as $\lambda_s$ reaches a maximum just below the assumed freezing temperature $T_f$~$=$~12~K. A peak in the ZF $\mu$SR relaxation rate is commonly associated with spin freezing, which is in good agreement with the complete decoupling of the ZF relaxation in a relative modest longitudinal field of 0.02~T at $T$~$=$~1.4~K as illustrated in Fig.~\ref{Fig4}(b). It is common to observe a two-component ZF-$\mu$SR relaxation function in spin glasses \cite{02_wiebe, 03_wiebe}. Furthermore, the overall $T$-dependence of the two signal components is similar to the case of Ba$_2$YReO$_6$ \cite{10_aharen_2}, which has also been proposed to host a spin-glass ground state. Finally, it is interesting to note that $\mu$SR is insensitive to the short-range magnetic correlations of Sr$_2$YReO$_6$ over most of the temperature range between $T_f$ and $\sim$~50~K where they are detected by other techniques, therefore the fluctuations of these spin clusters must be predominantly characterized by a frequency outside the $\mu$SR time window in this regime.

\subsection{C. Sr$_2$InReO$_6$: Magnetic Properties}

\begin{figure*}
\centering
\scalebox{0.5}{\includegraphics{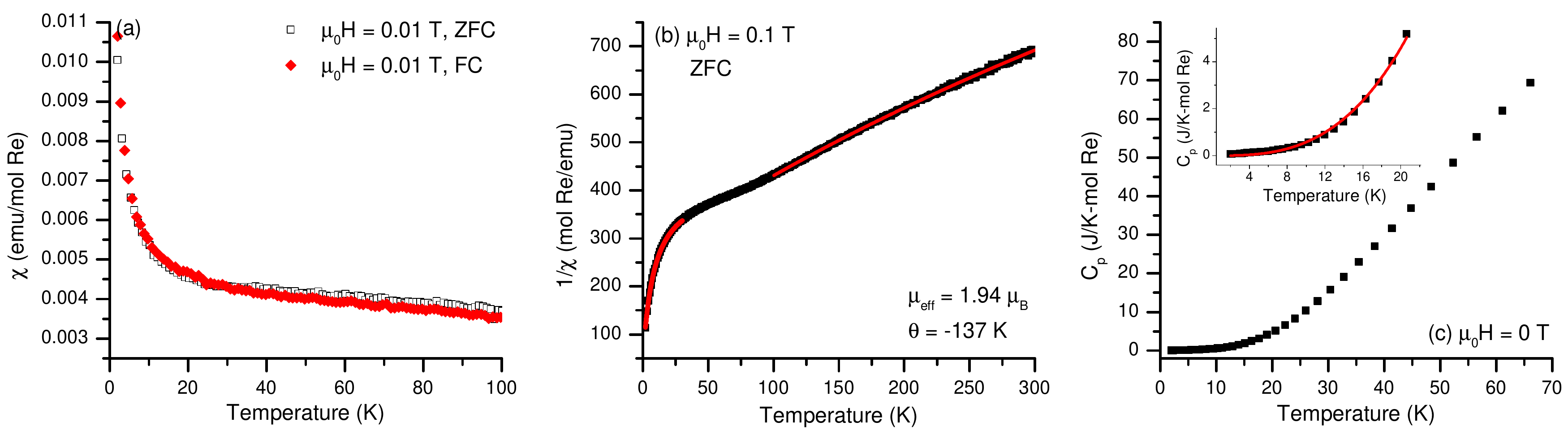}}
\caption{\label{Fig5} (a) DC magnetic susceptibility $\chi$ of Sr$_2$InReO$_6$ measured in an applied field of $\mu_0 H$~$=$~0.01~T. There is no evidence for magnetic ordering or ZFC/FC divergence down to 2~K. (b) The $\chi$ data ($\mu_0 H$~$=$~0.1~T) fits well to a Curie-Weiss law plus a constant for 100~$\le$~$T$~$\le$~300~K with an effective moment $\mu_{eff}$~$=$~1.94~$\mu_B$ and a Curie-Weiss temperature of -137~K. The low $T$ data between 2 and 30 K for $\mu_0 H$~$=$~0.1~T is also well-described by a Curie-Weiss law plus a constant, albeit with a drastically-reduced effective moment corresponding to only $\sim$~4\% of the total Re$^{5+}$ moments. (c) The heat capacity $C_p$ of Sr$_2$InReO$_6$ for $\mu_0 H$~$=$~0 T. No obvious magnetic features are visible in this data. The inset shows the low-$T$ $C_p$ data, with the fit described in the text illustrated by a solid curve.}
\end{figure*}

Figure~\ref{Fig5}(a) depicts the DC magnetic susceptibility $\chi$ measured in an applied field of $\mu_0 H$~$=$~0.01~T. There is no evidence for magnetic ordering or ZFC/FC divergence down to 2~K; the latter is in contrast to previous work which reports the onset of ZFC/FC divergence around 250~K \cite{11_gao}. While we cannot rectify the difference between our $\chi$ data and the earlier work, we repeated the $\chi$ measurements on our samples multiple times and found that all of the datasets were consistent with one another. Our high temperature data with $\mu_0 H$~$=$~0.1~T is well-described by a Curie-Weiss law plus a constant, with a fit between 100 and 300~K yielding $\theta_{CW}$~$=$~-137~K and an effective moment $\mu_{eff}$~$=$~1.94~$\mu_B$. While the effective moment has a similar value to Sr$_2$YReO$_6$, the Curie-Weiss temperature is significantly lower. There are deviations from the high temperature Curie-Weiss law with decreasing $T$, and in fact the 0.1~T data between 2 and 30~K fits well to a Curie-Weiss law plus a constant with significantly modified $\mu_{eff}$ and $\theta_{CW}$ values. The best fit in the low $T$ regime yields $\mu_{eff}$~$=$~0.38~$\mu_B$ and $\theta_{CW}$~$=$~-0.9~K, indicating a significant loss of magnetic moment at low $T$. These results can be understood in the context of a singlet ground state with $\sim$~4\% free spins, possibly arising from the excess Re found in the NPD refinements described above.

The specific heat $C_p$ shown in Fig.~\ref{Fig5}(c) is consistent with this interpretation, as no $\lambda$ anomaly characteristic of long-range ordering is observed at any temperature. Broad features are not apparent in the $C_p$ data either, in contrast to observations for Sr$_2$YReO$_6$. Finally, although the low $T$ $C_p$ data ($T$~$\le$~20~K) still fits well to the function $C_p$~$=$~$\gamma T + \beta T^3$ with $\gamma$~$=$~0.667(3)~mJ/K$^2$-mol Re and $\beta$~$=$~0.57(1)~mJ/K$^4$-mol Re, the magnitude of $\gamma$ is significantly reduced compared to Sr$_2$YReO$_6$. This finding is suggestive that the magnetic ground states of Sr$_2$YReO$_6$ and Sr$_2$InReO$_6$ are fundamentally different, although it is important to note that $\beta$ and $\gamma$ are very sensitive to the exact choice of fitting range. If we extend the low-$T$ fit up to 35~K, then we obtain similar $\beta$ and $\gamma$ parameters as reported in Ref.~\cite{11_gao}. In general, our $C_p$ data is in excellent agreement with that work. 

\begin{figure}
\centering
\scalebox{0.32}{\includegraphics{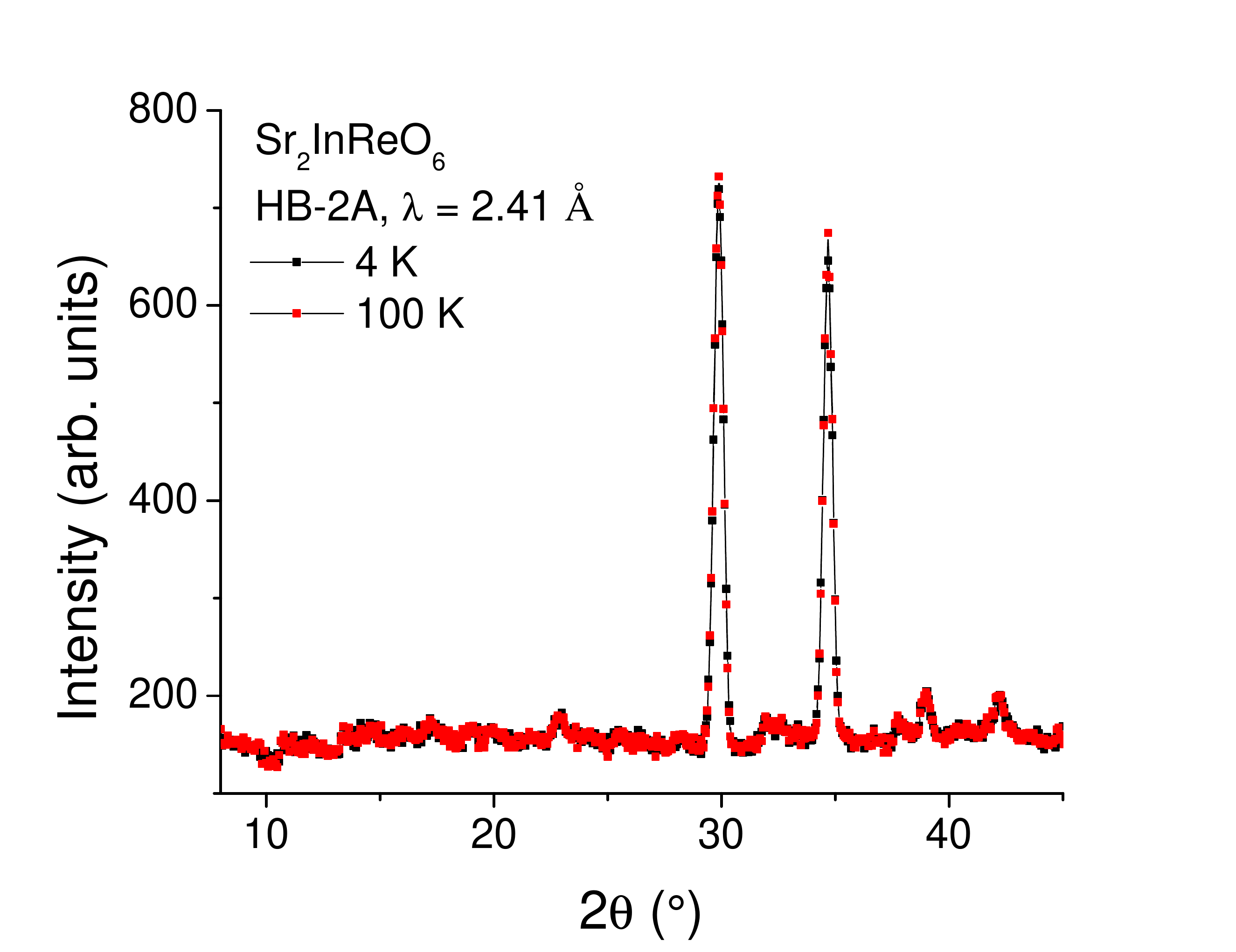}}
\caption{\label{Fig6} Neutron powder diffraction measurements with a wavelength of 2.41~\AA~at $T$~$=$~4 and 100~K for Sr$_2$InReO$_6$. }
\end{figure}

Neutron powder diffraction data with a wavelength of 2.41~\AA~for Sr$_2$InReO$_6$ is depicted in Fig.~\ref{Fig6} with $T$~$=$~4 and 100~K respectively. Similarly to the case of Sr$_2$YReO$_6$, the two diffraction patterns are nearly identical, and therefore there is no evidence for long-range magnetic order in this data either. The null diffraction result has similar implications to the Sr$_2$YReO$_6$ case: either the ordered Re moment is drastically-reduced as compared to its Ni counterparts or long-range order is absent down to 4~K. 

\begin{figure}
\centering
\scalebox{0.47}{\includegraphics{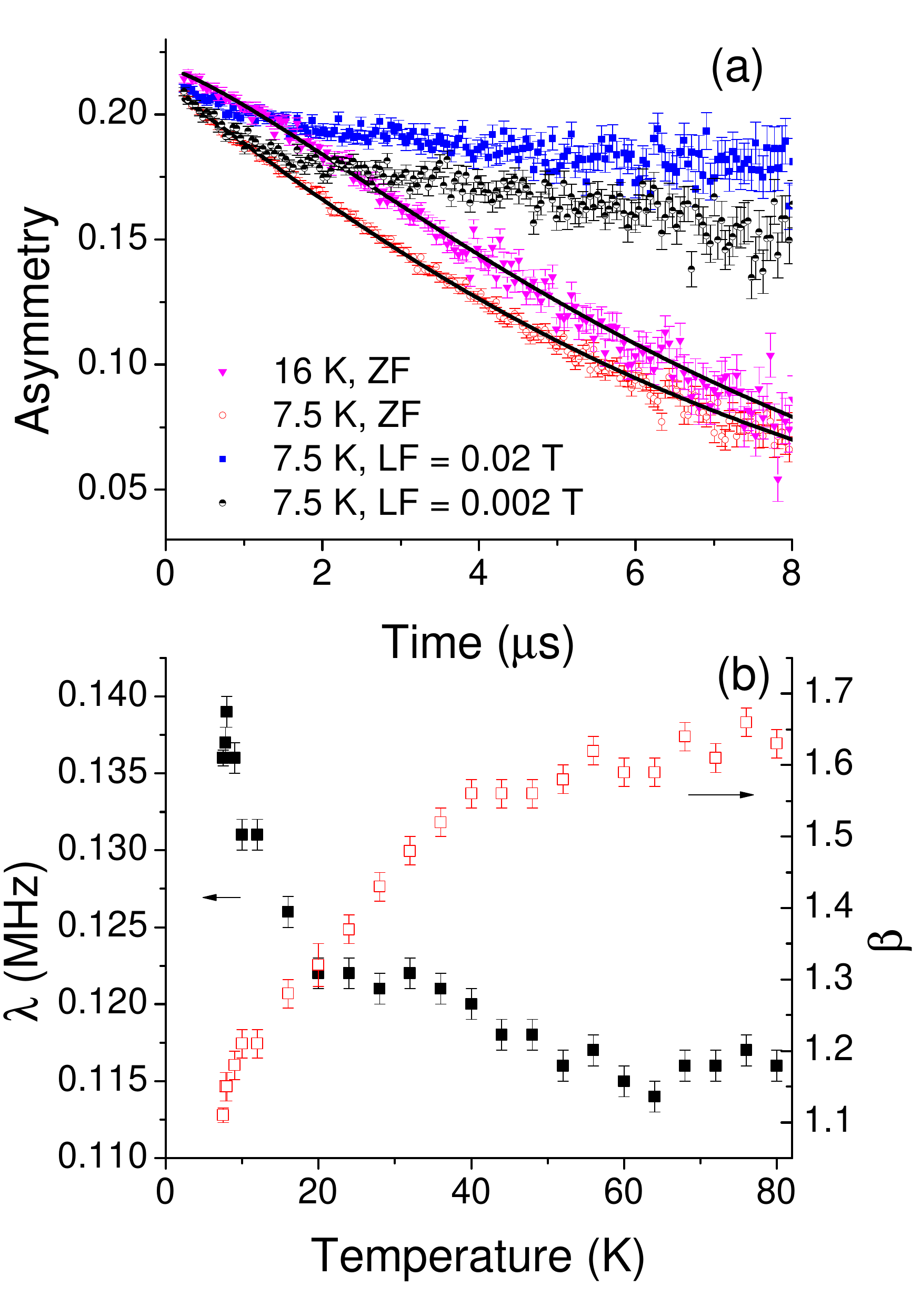}}
\caption{\label{Fig7} (a) ZF and LF-$\mu$SR spectra for Sr$_2$InReO$_6$ at selected temperatures. Most of the observed relaxation is due to static nuclear moments, although a very small quasi-static electronic component develops at low $T$. (b) $T$-dependence of the relaxation rate $\lambda$ and power $\beta$ for the ZF-$\mu$SR spectra.  }
\end{figure}

$\mu$SR proved to be useful here also for differentiating between these possibilities. Figure~\ref{Fig7} summarizes $\mu$SR results collected with the EMU spectrometer at ISIS. Figure~\ref{Fig7}(a) shows ZF-$\mu$SR asymmetry spectra at $T$~$=$~7.5 and 16~K. No missing asymmetry or spontaneous oscillations, as would be expected for a long-range ordered ground state, are apparent in the data. Instead, the spectra shows weak relaxation at all temperatures and can be fit well to the following function:
\begin{equation}
A(t) = A e^{-(\lambda t)^\beta}  
\end{equation}
The fitting results for $\lambda$ and $\beta$ are shown in Fig.~\ref{Fig7}(b). At the highest temperatures measured $\beta$~$\rightarrow$~2, which is consistent with expectations that the dominant relaxation mechanism in this regime arises from static nuclear moments. With decreasing $T$, $\beta$~$\rightarrow$~1 and $\lambda$ shows a small monotonic increase; this is precisely the behavior expected for a dilute, quasi-static electron spin system \cite{97_dereotier}. The LF-$\mu$SR spectra shown in Fig.~\ref{Fig7}(a) are in good agreement with this interpretation, as the ZF relaxation at $T$~$=$~7.5~K is significantly decoupled even in a modest applied LF of 0.002~T. Overall, the $\mu$SR data fully supports the picture of a singlet ground state coexisting with a small fraction of quasi-static free spins, suggested earlier on the basis of the $\chi$ measurements. We note that our $\mu$SR data for Sr$_2$InReO$_6$ looks qualitatively similar to $\mu$SR data for other proposed singlet ground state DPs, including Ba$_2$YMoO$_6$ \cite{10_aharen, 10_devries} and SrLaMgReO$_6$ \cite{15_thompson}.

\section{IV. Discussion}

We have firmly established that Sr$_2$YReO$_6$ hosts a spin-glass ground state, while Sr$_2$InReO$_6$ is characterized by a non-magnetic singlet ground state instead. Here, we make an effort to understand how the magnetic ground states of $d^2$ DPs evolve as a function of various tuning parameters, including spin-orbit coupling, structural distortions, and exchange interactions. There have now been several experimental reports in the literature describing the magnetic properties of particular members of this family, but an overall unifying picture is lacking. There are two different types of $d^2$ DPs, which are based on Os$^{6+}$ or Re$^{5+}$. There are no known $d^2$ DPs consisting of 3$d$ or 4$d$ transition metals, so the closest available comparison arises from considering Ni$^{2+}$~$S$~$=$~1 systems. Table~III summarizes several properties of various 5$d^2$ and Ni$^{2+}$~$S$~$=$~1 DPs, including the space group SpG, the tolerance factor $t$, the Curie-Weiss temperature $\theta_{CW}$, the ordering or freezing temperature $T_{c/f}$, and the magnetic ground state. 

Some trends in Table~III are readily apparent. Long-range ordered ground states are found for all of the Ni$^{2+}$ DP compounds previously investigated, with the type of long-range order determined by the signs and relative magnitudes of nearest neighbor (NN) exchange $J_1$ and next-nearest neighbor (NNN) exchange $J_2$. On the other hand, the magnetic ground states for the 5$d$ transition metal-based compounds are extremely varied. The 5$d$ systems with higher crystal symmetry show an increased tendency to exhibit magnetic order, while the lower symmetry systems are generally characterized by spin-glass or non-magnetic singlet ground states. On the basis of these observations alone, one obvious conclusion is that introducing enhanced spin-orbit coupling combined with significant structural distortions away from the ideal fcc lattice leads to a destabilization of magnetic order in this family. This finding is somewhat counterintuitive, as the geometric frustration generally argued to be responsible for the lack of long-range magnetic order in $d^2$ DPs should be maximized in cubic systems. 

A detailed understanding of the single ion ground state for 5$d^2$ systems is essential to understand this puzzling trend. The 5$d^2$ ions of cubic DPs are in an ideal octahedral oxygen environment, which leads to a very large $t_{2g}$/$e_g$ splitting $\Delta$. To satisfy Hund's first rule, the total spin for the two electrons is $S$~$=$~1. The orbital triplet degree of freedom must also be taken into account to determine the single ion ground state for a $d^2$ system, as orbital angular momentum $L$ is generally not quenched in 5$d$ magnets. If Hund's coupling $J_H$~$\ge$~2$\lambda_{SO}$, one finds that spin and orbital moments lock together to create a $J$~$=$~2 ground state \cite{11_chen, 13_plumb, 13_cook}.

In lower symmetry 5$d^2$ DPs, the five-fold degeneracy of the $J$~$=$~2 multiplet can be removed by a non-cubic crystal field term $\delta$, which is defined here as the energy difference between the ground state and the lowest-lying excited state. A non-magnetic singlet ground state ($J$~$=$~0) is even possible in principle for a significant $\delta$, due to the integer value for the angular momentum of the cubic crystal field ground state. In fact, recent RIXS work on Na$_2$IrO$_3$ \cite{13_gretarsson} and Sr$_3$CuIrO$_6$ \cite{12_liu} has proven that even modest structural distortions leading to non-cubic crystal fields at the Ir$^{4+}$ sites can induce large intramultiplet splitting on the order of 100's of meVs in 5$d$ magnets. Furthermore, non-cubic crystal field splitting has been argued to play an important role in stabilizing the ordered magnetic ground state for the 5$d^4$ DP Sr$_2$YIrO$_6$ \cite{14_cao}, since a non-magnetic single ion singlet state is expected for a cubic $d^4$ system in the strong SOC limit. 

\begin{center}
\begin{table}[htb]
\caption{Structural and magnetic properties of selected 3$d$ $S$~$=$~1 and 5$d^2$ double perovskite systems with a single magnetic sublattice} 

\begin{tabular}{| c | c | c | c | c | c | c |}
\hline
Compound & SpG & $t$ & $\theta_{CW}$ & $T_{c/f}$ & Ground state & Ref. \\  \hline
Ba$_2$WNiO$_6$ & {\it Fm$\bar{3}$m} & 1.05 & ? & ? & Type-II AF & \cite{67_cox}  \\
Sr$_2$WNiO$_6$ & {\it I4/m} & 0.99 & -175 & 54 & AFLRO & \cite{00_iwanaga} \\  
Sr$_2$TeNiO$_6$ & {\it C2/m} & 1 & -240 & 28-35 & AFLRO & \cite{00_iwanaga} \\ 
La$_2$TiNiO$_6$ & {\it P2$_1$/n} & 0.96 & -130 & 25 & Type-II AF & \cite{02_rodriguez} \\ 
SrLaSbNiO$_6$ & {\it P2$_1$/n} & 0.97 & ? & 26 & Type-I AF & \cite{92_attfield} \\
Ca$_2$WNiO$_6$ & {\it P2$_1$/n} &0.95 & -100 & 50 & Type-II AF & \cite{08_chakraborty} \\
Ba$_2$CaOsO$_6$ & {\it Fm$\bar{3}$m} & 0.99 & -156 & 50 & AFLRO & \cite{14_thompson} \\ 
Sr$_2$MgOsO$_6$ & {\it I4/m} & 1 & -347 & 110 & AFLRO & \cite{15_yuan} \\
Ca$_2$MgOsO$_6$ & {\it P2$_1$/n} & 0.96 & -72 & 19 & SG & \cite{15_yuan} \\
Ca$_3$OsO$_6$ & {\it P2$_1$/n} & 0.9 & -151 & 50 & NO LRO & \cite{13_feng} \\
Ba$_2$YReO$_6$ & {\it Fm$\bar{3}$m} & 1 & -616 & 50 & SG & \cite{10_aharen_2} \\
Sr$_2$YReO$_6$ & {\it P$2_1$/n} & 0.94 & -382 & 12 & SG & this work \\
La$_2$LiReO$_6$ & {\it P2$_1$/n} & 0.95 & -204 & 50 & SG/SS? & \cite{10_aharen_2} \\
SrLaMgReO$_6$ & {\it P2$_1$/n} & 0.97 & -161 & - & SS & \cite{15_thompson} \\ 
Sr$_2$InReO$_6$ & {\it P$2_1$/n} & 0.97 & -137 & - & SS & this work \\ \hline
\end{tabular}
The acronyms for the magnetic ground states specified in the table mean the following: AF (antiferromagnet), AFLRO (unknown antiferromagnetic long-range order), NO LRO (no long-range order detected), SG (spin glass), SS (non-magnetic singlet state). $\theta_{CW}$ and $T_{c/f}$ are both presented in K.
\end{table}
\end{center}

The enhanced sensitivity of 5$d$ systems to small structural distortions, as compared to their 3$d$ counterparts, arises from the spatially-extended nature of the 5$d$ orbitals. We propose here that this difference is one of the key factors leading to a wide variation of magnetic ground states for 5$d^2$ DPs. If $\delta$~$=$~0, which is valid for cubic systems, then the five-fold degeneracy of the $J$~$=$~2 multiplet will be preserved and a long-range ordered magnetic ground state will be the most common outcome \cite{11_chen}. On the other hand, if the intramultiplet splitting $\delta$ for the non-cubic systems becomes significant and creates a singlet ground state, there are three main possibilities based on the strength of the exchange interactions $J_{ex}$ between the magnetic ions. It is well-known that there is a critical value $x_c$~$=$~($J_{ex}$/$\delta$)$_c$ above which magnetic moments are induced in singlet systems \cite{08_goremychkin} and typical long-range magnetic order can still arise. On the other hand, for $x_c$~$>$~$J_{ex}$/$\delta$, spontaneous magnetism becomes impossible and therefore non-magnetic singlet ground states are realized instead. The third case with $x_c$~$\approx$~$J_{ex}$/$\delta$ proposed for the spin glass PrAu$_2$Si$_2$ is perhaps the most interesting, as it has been suggested that the spin freezing results from dynamic fluctuations of the crystal-field levels that destabilize the induced moments and frustrate the development of long-range magnetic order \cite{08_goremychkin}. This mechanism can serve as the source of both the frustration and disorder that are generally accepted to be essential ingredients in all spin glasses, and it can provide a natural explanation for why no obvious source of disorder has ever been found in 5$d^2$ spin glass DPs. 

Magnetic characterization studies of 5$d^2$ DPs yield results that are broadly consistent with our explanation given above for magnetic ground state selection. We first consider Sr$_2$YReO$_6$ and Sr$_2$InReO$_6$, which were extensively characterized in the present work. The Re$^{5+}$ sites of both systems are subjected to non-cubic crystal fields; this deviation from ideal fcc lattice geometry simultaneously enhances $\delta$ and reduces $J_{ex}$. One way to roughly quantify the former effect is by considering Re-O bond lengths. At $T$~$=$~4~K, our NPD measurements yield three inequivalent Re-O bond lengths of 1.95, 1.96, and 1.98~\AA~and 1.94, 1.97, and 2.01~\AA~for the Y and In systems respectively. The latter effect can also be quantified by considering the magnitude of the Curie-Weiss temperatures $\theta_{CW}$. Based on the significantly reduced value of $\theta_{CW}$ coupled with the enhanced structural distortion for Sr$_2$InReO$_6$ relative to Sr$_2$YReO$_6$, it is not surprising for a $J$~$=$~0 single ion ground state to arise for the former but not for the latter. 

In general, Table III indicates that 5$d^2$ systems with low Curie-Weiss temperatures and significant structural distortions tend to exhibit singlet ground states, while cubic systems show an increased tendency for magnetic order. Ba$_2$YReO$_6$ is the one known example of a cubic 5$d^2$ system that does not show long-range order, although broad, magnetic Bragg peaks were recently observed that are characteristic of Type I AF short-range order or a cluster spin glass \cite{thompson_private}. Furthermore, it is possible that Ba$_2$YReO$_6$ is not an ideal fcc system but rather exhibits a small structural distortion that can best be observed by synchrotron x-ray diffraction. 

Finally, we note that our discussion for 5$d^2$ systems does not extend to 4$d^1$/5$d^1$ systems since they cannot have single ion singlet ground states according to Kramer's theorem. On the basis of enhanced quantum fluctuations and strong geometric frustration, theoretical work for $d^1$ systems predicts that exotic collective spin-singlet ground states including quantum spin liquids and valence bond solids/glasses can be realized \cite{10_chen}. While we cannot rule out similar origins for the singlet ground states in 5$d^2$ DPs, previous theoretical work on these systems indicates that simple ordered states dominate the phase diagram instead \cite{11_chen}. 

\section{V. Conclusions}

We have performed detailed characterization studies of the 5$d^2$ double perovskites Sr$_2$InReO$_6$ and Sr$_2$YReO$_6$. Both systems crystallize in the space group {\it P$2_1$/n} and therefore non-cubic crystal fields are present at the Re$^{5+}$ sites. Our combined magnetic susceptibility, heat capacity, neutron powder diffraction, and muon spin relaxation measurements suggest that Sr$_2$YReO$_6$ is a spin glass while Sr$_2$InReO$_6$ is best characterized by a non-magnetic singlet ground state. We attribute the great diversity of magnetic ground states found for the 5$d^2$ double perovskite family, including the different ground states for Sr$_2$YReO$_6$ and Sr$_2$InReO$_6$ determined in this work, to a subtle interplay between spin-orbit coupling, non-cubic crystal fields, and exchange interactions. To verify our hypothesis for ground state selection in 5$d^2$ systems, it will be important to systematically determine the crystal field level schemes and exchange interactions for several members of this family.      


\begin{acknowledgments}
We thank C.M. Thompson and M.A. McGuire for useful discussions. This research was supported by the US Department of Energy (DOE), Office of Basic Energy Sciences. A.A.A. and S.C. were supported by the Scientific User Facilities Division, and J.-Q.Y. was supported by the Materials Science and Engineering Division. Z.Y.Z. acknowledges the CEM, and NSF MRSEC, under grant DMR-1420451. The neutron diffraction experiments were performed at the High Flux Isotope Reactor, which is sponsored by the Scientific User Facilities Division.  
\end{acknowledgments}

\end{document}